\documentstyle[12pt,aps]{revtex}%

\newcommand{\bce}{\begin{center}}
\newcommand{\ece}{\end{center}}
\newcommand{\be}{\begin{equation}}
\newcommand{\ee}{\end{equation}}
\newcommand{\bea}{\vspace{0.25cm}\begin{eqnarray}}
\newcommand{\eea}{\end{eqnarray}}

\def\lsim{\mathrel{\rlap{\lower4pt\hbox{\hskip1pt$\sim$}}
    \raise1pt\hbox{$<$}}}         
\def\gsim{\mathrel{\rlap{\lower4pt\hbox{\hskip1pt$\sim$}}
    \raise1pt\hbox{$>$}}}         

\def\PLA{{Phys. Lett.}  A }

\def\PRA{{Phys. Rev.} A }

\begin{document}

\title{{\LARGE {\bf A first test of Wigner function local realistic model}}}

\author{G.Brida, M.Genovese \footnote{ genovese@ien.it. Tel. 39 011 3919253, fax 39 011 3919259}, M. Gramegna, C.Novero}
\address{Istituto Elettrotecnico Nazionale Galileo Ferraris, Str. delle Cacce\\
91,I-10135 Torino, }
\author{E. Predazzi}
\address{Dip. Fisica Teorica Univ. Torino and INFN, via P. Giuria 1, I-10125 Torino. }
\maketitle

\vskip 1cm
{\bf Abstract}
\vskip 0.5cm 
We perform a first experimental test of the local realistic model of Ref. \cite{Santos2}. Our results disfavour the model and confirm standard quantum mechanics predictions. 
\vskip 1.5cm



\vspace{8mm}

In 1935 Einstein-Podolsky-Rosen \cite{EPR} suggested that Quantum Mechanics (QM) could be an incomplete theory, representing a statistical approximation of a complete deterministic theory where observable values are fixed by some hidden variable. A fundamental progress in discussing possible extensions of QM was the discovery of Bell \cite{Bell} that any realistic Local Hidden Variable (LHV) theory must satisfy certain inequalities, which can be violated in QM leading  to the possibility of an experimental test of the validity of QM as compared to LHV. Since then many experiments have been devoted to a test of Bell inequalities \cite{Mandel,asp,franson,type1,type2,ou,Kwiat,nos}, leading to a substantial agreement with quantum mechanics and disfavouring realistic local hidden variable theories.
However, due to the low total detection efficiency, no experiment has yet been able to exclude definitively realistic
local hidden variable theories \footnote{ It must be noticed that a recent experiment \cite{Win} performed using Be ions has reached very high efficiencies, around 98 \%, but in this case the two subsystems (the two ions) are not really separated systems during measurement and the test cannot be considered a real implementation of a detection loophole free test of Bell inequalities \cite{Vai}.}, for it is necessary a further additional
hypothesis \cite{santos}, stating that the observed sample of particles
pairs is a faithful subsample of the whole. This problem is known as detection or efficiency loophole. Because of this loophole a local realistic model can still be built in order to reproduce present experimental data.

Very recently Casado et al. \cite{Santos2} have presented a local realistic model addressed to be compatible with the available experimental data on Bell inequalities obtained with entangled photons, which are the most severe test of local realism realised up to now. This model represents the completion of series of papers where this scheme has been developed \cite{Santosv}. 
The main idea is that the probability distribution for the hidden variable is given by the Wigner function, which is positive for photons experiments. Furthermore a model of photodetection, which departs from quantum theory,  is built in order to reproduce available experimental results.

A great merit of this model is that it gives a number of constraints, which do not follow from the quantum theory and are experimentally testable. 

In particular, there is a minimal light signal level which may be reliably detected: a difference from quantum theory is predicted at low detection rates, namely when the single detection rate $R_S $ is lower than

\begin{equation}
R_S < { \eta F^2 R_c^2 \over 2 L d^2 \lambda \sqrt{ \tau T} }
\label{rate}
\end{equation}
where $\eta$ is the detection quantum efficiency, F is the focal distance of the lens in front of detectors, $R_c$ is the radius of the active area of the non-linear medium where entangled photons are generated, $\tau$ is the coherence time of incident photons, d is the distance between the non-linear medium and the photo-detectors, $\lambda$ the average wavelength of detected photons. L and T are two free parameters which are less well determined by the theory \cite{Santos3}: L can be interpreted as the active depth of the detector, while T is the time needed for the photon to be absorbed and should be approximately less than 10 ns \cite{Santos3}, being, in a first approximation, the length of the wave packet divided for the velocity of light.

How the departure from quantum mechanics would happen is not completely clear in Ref.\cite{Santos2}, anyway a lowering of the visibility should be expected \cite{Santos3}.

In order to perform a first experimental test of this model we have reconsidered  our set-up for measuring Bell inequalities based on the generation of non-maximally entangled photon states by the superposition of the parametric down conversion (PDC) light of two type I crystals \cite{nos}.       
The use of non-maximally entangled state was suggested by
the theoretical discovery that for non-maximally entangled pairs a total
efficiency larger than 0.67 \cite{eb} (in the limit of no background) is required to obtain an efficiency-loophole free experiment (even if the violation turns to be smaller), whilst for maximally
entangled pairs this limit rises to 0.81. 

More in details, our experiment follows and develops \cite{napoli} an idea by Hardy \cite
{hardy} and contemplates the creation of a polarisation (non maximally-)
entangled states of the form

\begin{equation}
\vert \psi \rangle = {\frac{ \vert H \rangle \vert H \rangle + f \vert V
\rangle \vert V \rangle }{\sqrt {(1 + |f|^2)}}}  \label{Psi}
\end{equation}
(where $H$ and $V$ indicate horizontal and vertical polarisations
respectively). 
The state of Eq. \ref{Psi} would be maximally entangled when $f=1$. 

With reference to the sketch of figure 1, our set up is made of two crystals of LiIO$_3$ (10x10x10 mm) 250 mm apart, a distance smaller than the coherence length of the pumping laser. This guarantees undistinguishability in the creation of a couple of photons in the first or in the second crystal. A couple of planoconvex lenses of 120 mm focal length centred in between focalises the spontaneous emission from the first crystal into the second one maintaining the angular spread. A hole of 4 mm diameter is drilled into the centre of the lenses in order to allow transmission of the pump radiation without absorption and, even more important, without adding stray-light, because of fluorescence and diffusion of the UV radiation.   This configuration, which performs as a so-called "optical condenser", was chosen among others as a compromise between minimisation of aberrations (mainly spherical and chromatic) and losses due to the number of optical components. The pumping beam at the exit of the first crystal is displaced from its input direction by birefringence: the small quartz plate (5 x5 x5 mm) in front of the first lens of the condensers compensates this displacement, so that the input conditions are prepared to be the same for the two crystals. Finally, a half-wavelength plate immediately after the condenser rotates the polarisation of the Argon beam and excites in the second crystal a spontaneous emission cross-polarised with respect to the first one. The dimensions and positions of both plates are carefully chosen in order not to intersect the spontaneous emissions at 633 and 789 nm. With a phase matching angle of $51^o$ they are emitted at $3.5^o$ and $4^o$ respectively. 

We use as photo-detectors two avalanche photodiodes with active quenching (EG\&G SPCM-AQ) with a sensitive area of 0.025 $mm^2$ and dark count below 50 counts/s.
PDC signal is coupled to an optical fibre (carrying the light on  the detectors) by means of a microscope objective with magnification 20, preceded by a polariser.

The output signals from the detectors are routed  to a two channel counter, in order to have the number of events on single channel, and to a  
Time to Amplitude Converter circuit, followed by a single channel analyser, for
selecting and counting coincidence events.

In our experiment the single detection rate is at the most of $R_S = 10^5$ counts per second.
Referring to the parameters of Eq. \ref{rate} we have $\eta = 0.51 \pm 0.02$ (a value which we have directly measured by using PDC detector calibration \cite{JMO}), F= 0.9 cm, $R_c = 1 $ mm, d= 0.75 m and $\tau = 4.2 \cdot 10^{-13} s$ (due to spectral selection by an interferential filter). L can be estimated of $3 \cdot 10^{-5}$ m \cite{L}.
This leads to $T \gsim 1 s$, extremely higher than the limit of 10 ns suggested in the model \cite{Santos2}. Thus, we are strictly in the condition where quantum mechanics predictions are expected to be violated and, in particular, a strong reduction of visibility is expected.

Nevertheless, our experimental data on coincidences of the two photodetectors show, when the angle of a polariser is varied, a high visibility, $V=0.98 \pm 0.01$.
This result is in  strong disagreement with the Wigner function local realistic model of ref. \cite{Santos2}.

With the purpose of performing a further test of model let us consider
our results about the Clauser-Horne sum:

\begin{equation}
CH=N(\theta _{1},\theta _{2})-N(\theta _{1},\theta _{2}^{\prime })+N(\theta
_{1}^{\prime },\theta _{2})+N(\theta _{1}^{\prime },\theta _{2}^{\prime
})-N(\theta _{1}^{\prime })-N(\theta _{2})  \label{eq:CH}
\end{equation}
which is strictly negative for every local realistic theory. In (\ref{eq:CH}), $N(\theta _{1},\theta _{2})$ is the number of coincidences between
channels 1 and 2 when the two polarisers are rotated to an angle $\theta _{1}$
and $\theta _{2}$ respectively, whilst $N(\theta _{1}^{\prime } )$ and $N(\theta _{2})$ denote single counts on detectors 1 and 2 respectively.

For quantum mechanics $CH$ can be larger than zero. For a maximally
entangled state the largest value is obtained for $\theta _{1}=67^{o}.5$ , $%
\theta _{2}=45^{o}$, $\theta _{1}^{\prime }=22^{o}.5$ , $\theta _{2}^{\prime
}=0^{o}$. 

For non-maximally entangled states the angles for which CH is maximal are
somehow different and the largest value smaller. The angles corresponding to
the maximum can be evaluated maximising Eq. \ref{eq:CH} with 

\bea
\left. \begin{array}{l}

  N[\theta _{1},\theta _{2}] =  [ \epsilon _1^{||} \epsilon _2^{||} (Sin[\theta _{1}]^{2}\cdot Sin[\theta_{2}]^{2}) + \\
  \epsilon _1^{\perp} \epsilon _2^{\perp} 
(Cos[\theta _{1}]^{2} \cdot Cos[\theta _{2}]^{2} )\\
  (\epsilon _1^{\perp} \epsilon _2^{||} Sin[\theta _{1}]^2\cdot Cos[\theta _{2}]^2 + \epsilon _1^{||} \epsilon _2^{\perp} 
Cos[\theta _{1}]^2 \cdot Sin [\theta _{2}]^2 )  \\
 + |f|^{2}\ast (\epsilon _1^{\perp} \epsilon _2^{\perp} (Sin[\theta _{1}]^{2}\cdot Sin[\theta_{2}]^{2}) +  \epsilon _1^{||} \epsilon _2^{||}
(Cos[\theta _{1}]^{2} Cos[\theta _{2}]^{2} ) +\\
(\epsilon _1^{||} \epsilon _2^{\perp} Sin[\theta _{1}]^2\cdot Cos[\theta _{2}]^{2} +\\
 \epsilon _1^{\perp} \epsilon _2^{||} 
Cos[\theta _{1}]^2 \cdot Sin [\theta _{2}]^2 )   \\
  +  (f+f^{\ast }) ((\epsilon _1^{||} \epsilon _2^{||} + \epsilon _1^{\perp} \epsilon _2^{\perp} - \epsilon _1^{||} \epsilon _2^{\perp} - 
\epsilon _1^{\perp} \epsilon _2^{||})
\cdot (Sin[\theta _{1}]\cdot Sin[\theta _{2}]\cdot
Cos[\theta _{1}]\cdot Cos[\theta _{2}]) ]  /(1+|f|^{2}) \,
\end{array}\right. \, .   
\eea
where (for the case of non-ideal polariser) $\epsilon _i^{||}$ and $\epsilon _i^{\perp}$ 
correspond to the transmission when the polariser (on the branch $i$)  axis is aligned or normal the polarisation axis respectively.

In our experiment we have created  a state with $f \simeq 0.4$ which gives the largest violation of Clauser-Horn inequality for $\theta_1 =72^o.24$, $\theta_2=45^o$, $\theta_1 ^{\prime}= 17^o.76$ and  $\theta_2 ^{\prime}= 0^o$.

However, because of low detection efficiency we must, as in any experiment performed up to now, substitute in Eq. \ref{eq:CH}  single counts $N(\theta _{1}^{\prime } )$ and $N(\theta _{2})$ with coincidence counts
$N(\theta _{1}^{\prime },\infty )$ and $N(\infty ,\theta _{2})$, where
$\infty $ denotes the absence of selection
of polarisation for that channel. This is one of the form in which detection loophole manifests itself. In the following we will denote Clauser-Horn sum where the former substitution is done with  $CH_{exp}$.

Our experimental result is $CH_{exp} = 513 \pm 25$ coincidences per second, 
which is more than 20 standard deviations from zero and compatible with the theoretical value predicted by quantum mechanics. 
Thus, our results show  a strong violation of Clauser-Horn inequality, in  agreement with standard quantum mechanics, even in a regime where the bound of Eq. \ref{rate} is strongly violated  and therefore strongly disfavour the model of Ref. \cite{Santos2}.

When more strict bounds on the parameters of the model and  more precise experimental predictions will be available \cite{Santos3} an experiment, specifically addressed to testing it, will be able to give  a conclusive test of the model itself.

\bigskip 

\vskip1cm \noindent {\bf Acknowledgements} \vskip0.3cm We would like to
acknowledge support of MIUR.

\vskip 2cm

\centerline{\bf Figure caption}
- Sketch of the source of polarisation entangled photons. NLC1 and NLC2 are two $LiIO_3$ crystals cut at the phase-matching angle of $51^o$. L1 and L2 are two identical piano-convex lenses with a hole of 4 mm in the centre. C is a 5 x 5 x 5 mm quartz plate for birefringence compensation and $\lambda / 2$ is a first order half wave-length plate at 351 nm. UV identifies the pumping radiation at 351 nm. 
\end{document}